\documentclass[9pt,journal,a4paper,oneside,onecolumn]{IEEEtran}

\usepackage{geometry}
\usepackage{graphicx}
\usepackage{amsmath}

\makeatletter
\def\@endtheorem{\endtrivlist}
\makeatother
\newtheorem{Step}{Step}

\newcommand{\BibTeX}{{\rmfamily B\kern-.05em{\scshape i\kern-.025em
b}\kern-.08em T\kern-.1667em\lower.7ex\hbox{E}\kern-.125emX}}

\begin{document}

\title{A DNA Sequence Compression Algorithm Based on LUT and LZ77}
\author{\authorblockN{Sheng~Bao} ~\IEEEmembership{Student Member,~IEEE,} \\
\authorblockA{Dept. of Information Engineering,Nanjing Univ.of P.\&
T.,Nanjing 210046,CHINA  \\Email : forrestbao@yahoo.com.cn},\\
\authorblockN{Shi Chen} \\
\authorblockA{School of Life Science,Nanjing University,Nanjing
210093,CHINA  \\Email : gattacalab@gmail.com}\\
\authorblockN{Zhiqiang Jing} \\
\authorblockA{Gattaca Lab,School of Life Science,Nanjing
University,Nanjing 210093,CHINA  \\Email : jzq8255@sina.com} \\
\authorblockN{Ran Ren}\\
\authorblockA{Dept. of Telecommunication Engineering,Nanjing Univ. of P.
\& T.,Nanjing 210046,CHINA  \\Email : george.r.ren@gmail.com}}
\maketitle

%  \begin{flushright}
%  \textit{"Nature must be interpreted as matter, energy, and information"
% }\\
%  ---Jeremy Campbell, Grammatical Man
%  \end{flushright}

\begin{abstract}
This article introduces a new DNA sequence compression algorithm which is based on LUT and LZ77 algorithm. Combined a LUT-based pre-coding routine and LZ77 compression routine,this algorithm can approach a compression ratio\footnote{This compression rate is defined as the compressed file size divided by base number} of 1.9bits \slash base and even lower.The biggest advantage of this algorithm is fast execution, small memory occupation and easy implementation.
%After the pre-coding routine,LZ77 algorithm is applied onto the processed sequence to finish the compression job.
%We hope more detailed work scan make our algorithm better.
\end{abstract}


%Attention:Since we haven't read through enough references, maybe the same work has already been done by others before.If so, we will cancel our paper and apologize to the authors who devised this algorithm before us.

%\IEEEpeerreviewmaketitle

\section{Introduction}
As Cohen\cite{CS_BS} said on Communication of The ACM said,``Biologists are aware of the degree of difficulty in days, months, or years in validating a given conjecture by lab experiment. Computer scientists are sure to benefit from being active and assertive partners with biologists",DNA sequence compression has caught the attention of some computer and biology scientists.\cite{Repeat} \cite{ACM2005}

%The information theory indicates two approaches to increase coding efficiency,decreasing dependency between signals and making their appearing probability equal.
Many researches on DNA sequence compression are devised. Some of them use the property of repeat in DNA sequence.\cite{Repeat}We devised a new DNA sequence compression algorithm combining LZ77 and the pre-coding routine which maps the combination of ATGC into 64 ASCII characters.

Since the essence of compression is a mapping between source file and destination file,the compression algorithm dedicates to find the relationship. In EE design,some logical circuits are implemented by FPGA. The key idea of FPGA is LUT(Look-up Table).Every input signal looks up its corresponding output which is stored in the chip before. We migrate this idea to our research on DNA sequence compression. We are trying to build a finite LUT which implements the mapping relationship of our coding process.

Some experiments indicate that the compression ratio is 1.9bits$\slash$base and even better. \cite{Shannon1948} \cite{The_Theory_of_Information_and_Coding} \cite{Information_Theory_Coding_and_Cryptography} Our algorithm is just adding a pre-coding process before the LZ77 compression. Compared with other DNA compression algorithms,the biggest advantage is fast execution, small memory occupation and easy implementation.
% What is more, this algorithm could be used as a companion to the existing ones.\cite{ACM2005}

\section{pre-coding Routine}
\subsection{The Look-up Table}
\label{sectionLUT}
The look-up table describes a mapping relationship between DNA segment and its corresponding characters. We assume that every three characters in source DNA sequence(without N\footnote{N refers to those not available or unknown base in DNA sequence}) will be mapped
into a character chosen from the character set which consists of 64 ASCII characters.\cite{ASCII}The look-up table is various. You can choose other character sets for coding. The one that we used is given in table \ref{table}. The braces behind each character contains the corresponding ASCII codes of these characters. For easy implementation,characters a,t,g,c and A,T,G,C will no longer appear in pre-coded file.

For instance,if a segment ``ACTGTCGATGCC" has been read,in the destination file,we represent them as ``j2X6". Obviously,the destination file is case-sensitive.

\begin{table*}[tbp]
 \caption{Look-up Table that we used} \label{table}
\begin{center}
\begin{tabular}[!]{c @{(} c @{)}|c|  c @{(} c @{)}|c|  c @{(} c @{)}|c|  c @{(} c @{)}|c}
\hline \multicolumn{2}{c|}{char} & bases & \multicolumn{2}{|c|}{char} & bases & \multicolumn{2}{|c|}{char} & bases &\multicolumn{2}{|c|}{char} & bases \\
\hline 	\symbol{33}  & 33  & A  A  A & \symbol{98}  & 98  & A  A  T & \symbol{34}  & 34  & A  A  C & \symbol{100} & 100 & A  A  G \\
	\symbol{101} & 101 & A  T  A & \symbol{102} & 102 & A  T  T & \symbol{35}  & 35  & A  T  C & \symbol{104} & 104 & A  T  G \\
	\symbol{105} & 105 & A  C  A & \symbol{106} & 106 & A  C  T & \symbol{107} & 107 & A  C  C & \symbol{108} & 108 & A  C  G \\
	\symbol{109} & 109 & A  G  A & \symbol{110} & 110 & A  G  T & \symbol{111} & 111 & A  G  C & \symbol{112} & 112 & A  G  G \\
	\symbol{113} & 113 & T  A  A & \symbol{114} & 114 & T  A  T & \symbol{115} & 115 & T  A  C & \symbol{36}  & 36  & T  A  G \\
	\symbol{117} & 117 & T  T  A & \symbol{118} & 118 & T  T  T & \symbol{119} & 119 & T  T  C & \symbol{120} & 120 & T  T  G \\
	\symbol{121} & 121 & T  C  A & \symbol{122} & 122 & T  C  T & \symbol{37}  & 37  & T  C  C & \symbol{66}  & 66  & T  C  G \\
	\symbol{38}  & 38  & T  G  A & \symbol{68}  & 68  & T  G  T & \symbol{69}  & 69  & T  G  C & \symbol{70}  & 70  & T  G  G \\
	\symbol{39}  & 39  & C  A  A & \symbol{72}  & 72  & C  A  T & \symbol{73}  & 73  & C  A  C & \symbol{74}  & 74  & C  A  G \\
	\symbol{75}  & 75  & C  T  A & \symbol{76}  & 76  & C  T  T & \symbol{77}  & 77  & C  T  C & \symbol{78}  & 78  & C  T  G \\
	\symbol{79}  & 79  & C  C  A & \symbol{80}  & 80  & C  C  T & \symbol{81}  & 81  & C  C  C & \symbol{82}  & 82  & C  C  G \\
	\symbol{83}  & 83  & C  G  A & \symbol{40}  & 40  & C  G  T & \symbol{85}  & 85  & C  G  C & \symbol{86}  & 86  & C  G  G \\
	\symbol{87}  & 87  & G  A  A & \symbol{88}  & 88  & G  A  T & \symbol{89}  & 89  & G  A  C & \symbol{90}  & 90  & G  A  G \\
	\symbol{48}  & 48  & G  T  A & \symbol{49}  & 49  & G  T  T & \symbol{50}  & 50  & G  T  C & \symbol{51}  & 51  & G  T  G \\
	\symbol{52}  & 52  & G  C  A & \symbol{53}  & 53  & G  C  T & \symbol{54}  & 54  & G  C  C & \symbol{55}  & 55  & G  C  G \\
	\symbol{56}  & 56  & G  G  A & \symbol{57}  & 57  & G  G  T & \symbol{43}  & 43  & G  G  C & \symbol{45}  & 45  & G  G  G \\	\hline
\end{tabular}%
\end{center}
\end{table*}

%The LZ-77 algorithm is applied after the pre-encoding routine.

\subsection{Handling the N}
As what we mentioned above,the character N refers to the segment which is unknown. Our experience taught us that N doesn't appear singly. Usually,scores or hundreds of Ns appear together. It is necessary to consider this situation which happens frequently in sequences under investigation.

When we encounter a serial of successive Ns, our algorithm inserts two ``\symbol{47}" into destination file to identify the starting and end place of these successive Ns. There is a number which equals to the number of Ns between the ``\symbol{47}" pair. For instance,if segment "NNNNNN" has been read,in the destination file,we represent them as ``\symbol{47}6\symbol{47}".

\subsection{Segment which consists of less than 3 non-N bases}
Non-N base is defined as the base which is not N. Thus,they are A,T,G or C. In section \ref{sectionLUT},we read bases 3 by 3.But in some conditions,we can't read three successive non-N bases. For example,the segment ``TCN" has been read. As what we mentioned,we handle ``N" differently as non-N bases. Then how do you process the segment ``TC"? You cannot find any arrangement in table \ref{table} which is ``TC".
In this circumstance,we just write the original segment into destination file.

\section{Algorithm steps}

%while(haven't reached the end of the source file)

This algorithm consists of two phases.The first one is from Step 1 to Step \ref{repeat} and the second one is the last step.

\begin{Step}
read 3 most beginning unprocessed characters.If successful,go along to step 2.Otherwise(the EOF \footnote{EOF means end-of-file which is the identifier of end of a file} is reached),process the last one or two characters by step \ref{directly}.
\end{Step}

\begin{Step}
judge whether there are all non-N characters.If it is,jump to step \ref{look}.Otherwise Process N by step \ref{N} and non-N bases by step \ref{directly} respectively.
\end{Step}

\begin{Step}
code the three characters according to their arrangement by table \ref{table} and write the coded character into destination file.
\label{look}
\end{Step}

\begin{Step}
read along successive Ns,write ``\symbol{47}n\symbol{47}" into destination file,where n is the number of successive Ns being processed.
After that,jump to step \ref{repeat}.
\label{N}
\end{Step}

\begin{Step}
write non-N characters whose number is less than three into destination file directly without any modification.
After that,jump to step \ref{repeat}.
\label{directly}
\end{Step}

\begin{Step}
Return to step 1 and repeat all process until EOF is reached.
\label{repeat}
\end{Step}

\begin{Step}
compress the output file by LZ77 algorithm \cite{LZ77} \cite{gzip}
\end{Step}

%These steps can be easily illustrated as: original file $\text{pre-coding routine} \atop {------- \to}$ pre-coded file $\text{LZ77} \atop {------- \to}$ final file

Here is an example illustrates how the algorithm works.``|" indicates different steps of the algorithm processing file.The segment in source file is

\centerline{ATG|CG|NNNNNNNNNNNNNNNNNNNN|ACC|GCC|ATC|TCT| CG|EOF}
The destination file of pre-coding routine is

\centerline{hCG/20/k6\symbol{35}zCG EOF}

The decompression routine is just the inverse operation of the compression routine. In \cite{glab},we introduce the most up-to-date C++
implementation of pre-coding algorithm,both coding and decoding ones.

\section{Algorithm Evaluation}
Since LZ77 has been proved to be accurate and efficient by earlier papers and user practice,we only consider the pre-coding routine.

\subsection{Accuracy}As to the DNA sequence storage, accuracy must be taken firstly in that even a single base mutation,insertion, deletion or SNP would result in huge change of phenotype as we see in the sicklemia. It is not tolerable that any mistake exists either in
compression or in decompression. Although not yet proved mathematically, it could be infer from table \ref{table} that our algorithm is accuracy,since every base arrangement uniquely corresponds to an ASCII character and all Ns will be mapped into a number braced by two ``\symbol{47}".

\subsection{Efficiency}
You can see that the pre-coding algorithm can compress original file from 3 characters into 1 character for any 3 non-N bases segment.
And destination file uses less ASCII character to represent successive Ns than source file,if the length of Ns is greater than 3. In practice,the length of Ns often is much greater than 3. The more Ns the source file contains,the more efficient the algorithm will be. So we can infer the file size becomes small.

\subsection{Time elapsed}
Today many compression algorithms are highly desirable, but they require considerable time to execute. Some of them take much more time than LZ77.In Chen's paper\cite{Repeat},compression of two sequences by some newly-developed algorithm elapsed more time than the traditional LZ77.As our algorithm is based on a LUT rather than sequence statistics, it can save the time of obtaining statistic information of sequence.

And more,after the pre-coding routine,the character number is $1/3$ of source one.The LZ77 will take less time to operate it than operating the source one.

Our experiments in section \ref{experiment} also provide proof to above conclusions.
You can see the elapsed time of our algorithm is in $10^{-3}$ second level whereas time elapsed of many newly-developed algorithm mentioned in Chen's paper\cite{Repeat} is in second level.\footnote{our CPU is only 50MHz faster than the CPU that they used.}Specially,the pre-coding routine only takes few $10^{5}$ CPU cycles which means it takes $10^{-4}$ second on a 1G CPU.

Not only faster than those newly-developed algorithm,our algorithm is also faster than LZ77,the algorithm used in many bioinformatics databases.Tabel \ref{time} indicates our algorithm takes almost only 26\% time of the one that Gzip needs.

\subsection{Space Occupation}
Our algorithm reads characters from source file and writes them immediately into destination file.It costs very small memory space to store only a few characters.The space occupation is in constant level.

In our experiments,the OS has no swap partition.All performance can be done in main memory which is only 256 MB on our PC.

\section{Compression Experiments}
\label{experiment}
Experiments are done to test our algorithm. Codes for testing the algorithm are continually revising.\cite{OurCode} \cite{glab} There should be some difference of experiment results between different versions of our codes.

These tests are performed on a computer whose CPU is AMD Duron 750MHz and OS is MagicLinux 1.2 (Linux Kernel 2.6.9) without swap partition. Testing programs are executed at multiuser text mode and compiled by gcc 3.3.2 without optimization. The file system where tests preformed is ext3 on a 4.3 GB Quantum Fireball hard disk with 5400 RPM(rounds per minute).Table \ref{size} lists the results in file size while table \ref{time} lists the cost of executing time.Appendix list all source codes of our algorithm and Linux shell by which we used Gzip to apply LZ77 algorithm.

\begin{table*}[ht]
\centering
\begin{tabular}{cccccc}
\hline
sequence 	& base number 	& file size(our algorithm) & file size(GZip) & $\overline{R}$(bits/base)\\ \hline
atatsgs 	& 9647 		& 18736 	& 20936 	& 1.9422 \\
atef1a23 	& 6022 		& 11448 	& 12272 	& 1.9010  \\
atrdnaf 	& 10014 	& 20256 	& 22816 	& 2.0228   \\
atrdnai 	& 5287 		& 10192 	& 9964 		& 1.9277   \\
chmpxx 		& 121024 	& 237744	& 276192 	& 1.9644   \\
chntxx 		& 155939 	& 309256 	& 364104 	& 1.9832   \\
hehcmvcg 	& 229354 	& 466296 	& 533888 	& 2.0331   \\
HSG6PDGEN 	& 52173 	& 102296 	& 117096 	& 1.9601   \\
HUMDYSTROP 	& 38770 	& 77504 	& 91624 	& 1.9991   \\
HUMHDABCD 	& 66495 	& 118424 	& 131848 	& 1.7809   \\
humghcsa 	& 58864		& 105864 	& 137344 	& 1.9683   \\
humhprtb 	& 56737 	& 112240 	& 128624 	& 1.9783   \\
mmzp3g 		& 10833 	& 21632 	& 25160 	& 1.9969   \\
mpomtcg 	& 186609 	& 377864 	& 434632 	& 2.0249   \\
mtpacg 		& 100314 	& 196760 	& 229944 	& 1.9614   \\
vaccg 		& 191737 	& 372688 	& 431792 	& 1.9437   \\
xlxfg512 	& 19338 	& 33864 	& 35408 	& 1.7512   \\ \hline
Average 	& -- 		& -- 		& -- 		& 1.9494   \\ \hline
\end{tabular}%
\caption{Experiment result on file size}
\label{size}
\end{table*}

\begin{table*}[ht]
\begin{center}
\begin{tabular}{cccccc}
\hline
sequence & second phase($10^{-3}$s) & pre-coding CLKs& decoding by LUT CLKs & Gzip($10^{-3}$s)\\ \hline
atatsgs & 8 & $<$10000 & $<$10000 & 13 \\
atef1a23 & 8  & $<$10000 & $<$10000 & 11 \\
atrdnaf & 9  & $<$10000 & $<$10000 & 14   \\
atrdnai & 9  & $<$10000 & $<$10000 & 10    \\
chmpxx & 23  & 30000 & 20000 & 105    \\
chntxx & 29  & 40000 & 20000 & 135    \\
hehcmvcg & 43  & 60000 & 30000 & 198    \\
HSG6PDGEN & 13  & 10000 & 10000 & 44    \\
HUMDYSTROP & 8  & 10000 & $<$10000 & 37    \\
HUMHDABCD & 14  & 10000 & 10000 & 50    \\
humghcsa & 14  & 20000 & 10000 & 55    \\
humhprtb & 14  & 10000 & 10000 & 49    \\
mmzp3g & 9  & $<$10000 & $<$10000 & 14    \\
mpomtcg & 34  & 40000 & 40000 & 100    \\
mtpacg & 20  & 20000 & 20000 & 88    \\
vaccg & 36  & 40000 & 30000 & 164    \\
xlxfg512 & 9  & $<$10000 & $<$10000 & 18    \\ \hline
Average & 17.647  & -- & -- & 65    \\ \hline
\end{tabular}%
\end{center}
\caption{Experiment result on executing time.}
\label{time}
\end{table*}
In table \ref{time},``second phase" refers to the sum of time elapsed in both compression and decompression process with error $10^{-3}$ second.``pre-coding CLKs" means the CPU clock cycle needed by compression in LUT pre-coding process while ``decoding by LUT CLKs" stands for the CPU clock cycle needed in decompressing the file coded by LUT.The CPU clock cycle is counted by every 10000 cycles. Considering nowadays computers are run $10^9$ cycles per second,the error is only $10^{-4}$ second.``Gzip" means the time needed by both compress and decompress the original sequence file.

In these experiments, we also have some new discoveries.

%The last step of our compression,the LZ77 compression, elapses more time than the pre-coding algorithm whose time complexity can be inferred as $O(n)$.
Many tests indicate that the
compression rate of only using LZ77 is almost the same as the one
of using pre-coding algorithm. The last step,applying LZ77 on to
the pre-coded file,only improves a little of compression rate.
It just compresses the file size into almost $75\%$ of the one
which is only compressed by pre-coding routine.

Considering the time cost by LZ77,the benefit is not so large. If
the time elapsed is very important,a wise choice is that the last
step should be skipped.

\section{Advantage of our algorithm}
Compared with other algorithms,the compression ratio of our algorithm is 0.2bits/base higher than theirs generally.
But the cost of the tiny 0.2 bits/base is very high.In Chen's paper\cite{Repeat},you can feel the time of program running since they are in second level.Some ones even cost minutes or hours of time to run.But our algorithm runs almost $10^{3}$ time faster than them.Compared with present LZ77 algorithm used widely in bioinformatics,our algorithm performances better than it in both compression ratio and elapsed time.
Our algorithm is very useful in database storing.You can keep sequences as records in database instead of maintaining them as files.By just using the pre-coding routine,users can obtain original sequences in a time that can't be felt.

Additionally,our algorithm can be easily implemented while some of them will take you more time to program.
\section{Conclusion}
In this article, we discussed a new DNA compression algorithm
whose key idea is LUT.It performance better than LZ77 in both compression ratio and elapsed time while the compression ratio is a little higher than newly-developed algorithms but many times faster than them. We are trying to do more work,such as combining our LUT pre-coding routine with other compression algorithms,to revise our algorithm
in order to improve its performance.

\section*{Acknowledgment}
Mr. Michael Shell in Dept. of Electrical and Computer Engineering of Georgia Institute of Technology,the author of IEEETran
\LaTeX  \ class and \BibTeX\ style package,gave us lots of instructions about using those two packages in our typesetting process.

Yao Liu from Dept. of Computer Science of Nanjing University wrote the decode program in pre-coding routine for us.Thanks for her help.

\bibliographystyle{IEEEtran}
\bibliography{lookuptable}

\appendices
\section{C++ implemetation of pre-coding process}
\begin{verbatim}
#include <iostream.h>
#include <fstream.h>
#include <string.h>
#include <stdlib.h>
#include <iomanip.h>
#include <math.h>
#define LENGTH 3


using namespace std;
using std::string;

char LUT(int a[]);//function used for pre-code regular segment

int main(void)
{

//define input file
ifstream infile;
string infilename;
cout<<"Enter the file name";
cin>>infilename;
infile.open(infilename.c_str());

if(!infile)
{
	cout<<"Can not open file"<<endl;
}
//end of defining input file 

//define output file
ofstream outfile;
string outfilename=infilename+".LUT";

outfile.open(outfilename.c_str());
//end of defining output file

int temp[LENGTH];//define temporary array for storing segment

char ch;
int i=0;//the number of bases in temporary array,minus 1
int count=0;//count the base number in original file
char enter;//code being writern into destination file
int numN=0;//number of Ns which are read

while(infile.get(ch))//omit the "ENTER",whose ASCII is 10
{
		if((ch!='N'||ch!='n')&& ch!=char(10))
		{
			if((int)ch>96)
			{
				int intch=(int)ch-32;
				ch=(char)intch;
			}//change small case char to large case char.a->A,t->T,c->C,g->G
			
			temp[i]=(int)ch;
			count++;
						
			if (i==2)
			{
				//cout<<temp[0]<<temp[1]<<temp[2]<<endl;
				enter=LUT(temp);
				//cout<<enter<<endl;
				outfile.put(enter);
				i=0;
			}
			else//haven't read three chars
			{	
				i++;
			}
		}
		if(ch=='N'||ch=='n')//situation of N
		{
			numN=1;
			while (infile.get(ch)&&(ch=='N'||ch=='n'))//if more N can be read
			{
				//cout<<"numN="<<numN<<endl;
				numN++;
			}
			outfile<<"/"<<numN<<"/";
			i=1;
			if((int)ch>96)
			{
				int intch=(int)ch-32;
				ch=(char)intch;
			}
			temp[0]=(int)ch;//the first non-N base will be stored in temp array
			//cout<<"first non-N"<<temp[0]<<endl;
		}
	//end of operate
}
//end of while
if(i!=0)
{
	i--;
	while(i!=-1)
	{
	enter=(char)temp[i];
	outfile.put(enter);
	i--;
	}
}
//output the last one or two chars


//cout<<"temp="<<temp[0]<<"--"<<temp[1]<<"--"<<temp[2]<<endl;
//cout<<"i="<<i<<endl;
infile.close();
outfile.close();
cout<<"There are totally "<<setw(4)<<count<<"character(s) in the original file"<<endl;
return 0;

}
//end of main function



char LUT(int temp[3])
{
char enter;
if(temp[0]!=78&&temp[1]!=78&&temp[2]!=78)
{
	if(temp[0]==65)//A
	{
		if(temp[1]==65){//A
			if(temp[2]==65){enter=(char)33;}  //A	
			else if(temp[2]==84){enter=(char)98;}  //T
			else if(temp[2]==67){enter=(char)34;}  //C
			else if(temp[2]==71){enter=(char)100;}  //G
		}
		if(temp[1]==84){//T
			if(temp[2]==65){enter=(char)101;}  //A
			else if(temp[2]==84){enter=(char)102;}  //T
			else if(temp[2]==67){enter=(char)35;}  //C
			else if(temp[2]==71){enter=(char)104;}  //G
		}
		if(temp[1]==67){//C
			if(temp[2]==65){enter=(char)105;}  //A
			else if(temp[2]==84){enter=(char)106;}  //T
			else if(temp[2]==67){enter=(char)107;}  //C
			else if(temp[2]==71){enter=(char)108;}  //G
		}
		if(temp[1]==71){//G
			if(temp[2]==65){enter=(char)109;}  //A
			else if(temp[2]==84){enter=(char)110;}  //T
			else if(temp[2]==67){enter=(char)111;}  //C
			else if(temp[2]==71){enter=(char)112;}  //G
		}
	}

	if(temp[0]==84)//T
	{
		if(temp[1]==65){//A
			if(temp[2]==65){enter=(char)113;}  //A
			else if(temp[2]==84){enter=(char)114;}  //T
			else if(temp[2]==67){enter=(char)115;}  //C
			else if(temp[2]==71){enter=(char)36;}  //G
		}
		if(temp[1]==84){//T
			if(temp[2]==65){enter=(char)117;}  //A
			else if(temp[2]==84){enter=(char)118;}  //T
			else if(temp[2]==67){enter=(char)119;}  //C
			else if(temp[2]==71){enter=(char)120;}  //G
		}
		if(temp[1]==67){//C
			if(temp[2]==65){enter=(char)121;}  //A
			else if(temp[2]==84){enter=(char)122;}  //T
			else if(temp[2]==67){enter=(char)37;}  //C
			else if(temp[2]==71){enter=(char)66;}  //G
		}
		if(temp[1]==71){//G
			if(temp[2]==65){enter=(char)38;}  //A
			else if(temp[2]==84){enter=(char)68;}  //T
			else if(temp[2]==67){enter=(char)69;}  //C
			else if(temp[2]==71){enter=(char)70;}  //G
		}
	}

	if(temp[0]==67)//C
	{
		if(temp[1]==65){//A
			if(temp[2]==65){enter=(char)39;}  //A
			else if(temp[2]==84){enter=(char)72;}  //T
			else if(temp[2]==67){enter=(char)73;}  //C
			else if(temp[2]==71){enter=(char)74;}  //G
		}
		if(temp[1]==84){//T
			if(temp[2]==65){enter=(char)75;}  //A
			else if(temp[2]==84){enter=(char)76;}  //T
			else if(temp[2]==67){enter=(char)77;}  //C
			else if(temp[2]==71){enter=(char)78;}  //G
		}
		if(temp[1]==67){//C
			if(temp[2]==65){enter=(char)79;}  //A
			else if(temp[2]==84){enter=(char)80;}  //T
			else if(temp[2]==67){enter=(char)81;}  //C
			else if(temp[2]==71){enter=(char)82;}  //G
		}
		if(temp[1]==71){//G
			if(temp[2]==65){enter=(char)83;}  //A
			else if(temp[2]==84){enter=(char)40;}  //T
			else if(temp[2]==67){enter=(char)85;}  //C
			else if(temp[2]==71){enter=(char)86;}  //G
		}
	}

	if(temp[0]==71)//G
	{
		if(temp[1]==65){//A
			if(temp[2]==65){enter=(char)87;}  //A
			else if(temp[2]==84){enter=(char)88;}  //T
			else if(temp[2]==67){enter=(char)89;}  //C
			else if(temp[2]==71){enter=(char)90;}  //G
		}
		if(temp[1]==84){//T
			if(temp[2]==65){enter=(char)48;}  //A
			else if(temp[2]==84){enter=(char)49;}  //T
			else if(temp[2]==67){enter=(char)50;}  //C
			else if(temp[2]==71){enter=(char)51;}  //G
		}
		if(temp[1]==67){//C
			if(temp[2]==65){enter=(char)52;}  //A
			else if(temp[2]==84){enter=(char)53;}  //T
			else if(temp[2]==67){enter=(char)54;}  //C
			else if(temp[2]==71){enter=(char)55;}  //G
		}
		if(temp[1]==71){//G
			if(temp[2]==65){enter=(char)56;}  //A
			else if(temp[2]==84){enter=(char)57;}  //T
			else if(temp[2]==67){enter=(char)43;}  //C
			else if(temp[2]==71){enter=(char)45;}  //G
		}
	}

}
return enter;
}
\end{verbatim}
\section{C++ implementation of decoding file coded by pre-coding process}
\begin{verbatim}
#include <iostream.h>
#include <fstream.h>
#include <string.h>
#include <stdlib.h>
#include <iomanip.h>
#include <math.h>
#define LENGTH 3

using namespace std;
using std::string;

int main(void)
{
    //define input file
    ifstream infile;
    string infilename;
    cout<<"Enter the file name";
    cin>>infilename;
    string outfilename=infilename+".de";
    //infilename=infilename+".LUT";
    infile.open(infilename.c_str());

    if(!infile)
    {
    	cout<<"Can not open file"<<endl;
    }
    //end of defining input file


    //define output file
    ofstream outfile;
    outfile.open(outfilename.c_str());
    //end of defining output file

    int i,j=0;
	char ch;
	if(!infile)
	{
		cout<<"can not open file"<<endl;
		return -1;
	}
	while(infile.get(ch))
	{
  		//infile.get(ch);
		if(ch!='/')   //in case that the char is not '/'
		{
			switch((int)ch)  //according to looking-up table
			{
				case 33:  outfile<<"AAA";break;
				case 98:  outfile<<"AAT";break;
				case 34:  outfile<<"AAC";break;
				case 100: outfile<<"AAG";break;
				case 101: outfile<<"ATA";break;
				case 102: outfile<<"ATT";break;
				case 35:  outfile<<"ATC";break;
				case 104: outfile<<"ATG";break;
				case 105: outfile<<"ACA";break;
				case 106: outfile<<"ACT";break;
				case 107: outfile<<"ACC";break;
				case 108: outfile<<"ACG";break;
				case 109: outfile<<"AGA";break;
				case 110: outfile<<"AGT";break;
				case 111: outfile<<"AGC";break;
				case 112: outfile<<"AGG";break;
				case 113: outfile<<"TAA";break;
				case 114: outfile<<"TAT";break;
				case 115: outfile<<"TAC";break;
				case 36:  outfile<<"TAG";break;
				case 117: outfile<<"TTA";break;
				case 118: outfile<<"TTT";break;
				case 119: outfile<<"TTC";break;
				case 120: outfile<<"TTG";break;
				case 121: outfile<<"TCA";break;
				case 122: outfile<<"TCT";break;
				case 37:  outfile<<"TCC";break;
				case 66:  outfile<<"TCG";break;
				case 38:  outfile<<"TGA";break;
				case 68:  outfile<<"TGT";break;
				case 69:  outfile<<"TGC";break;
				case 70:  outfile<<"TGG";break;
				case 39:  outfile<<"CAA";break;
				case 72:  outfile<<"CAT";break;
				case 73:  outfile<<"CAC";break;
				case 74:  outfile<<"CAG";break;
				case 75:  outfile<<"CTA";break;
				case 76:  outfile<<"CTT";break;
				case 77:  outfile<<"CTC";break;
				case 78:  outfile<<"CTG";break;
				case 79:  outfile<<"CCA";break;
				case 80:  outfile<<"CCT";break;
				case 81:  outfile<<"CCC";break;
				case 82:  outfile<<"CCG";break;
				case 83:  outfile<<"CGA";break;
				case 40:  outfile<<"CGT";break;
				case 85:  outfile<<"CGC";break;
				case 86:  outfile<<"CGG";break;
				case 87:  outfile<<"GAA";break;
				case 88:  outfile<<"GAT";break;
				case 89:  outfile<<"GAC";break;
				case 90:  outfile<<"GAG";break;
				case 48:  outfile<<"GTA";break;
				case 49:  outfile<<"GTT";break;
				case 50:  outfile<<"GTC";break;
				case 51:  outfile<<"GTG";break;
				case 52:  outfile<<"GCA";break;
				case 53:  outfile<<"GCT";break;
				case 54:  outfile<<"GCC";break;
				case 55:  outfile<<"GCG";break;
				case 56:  outfile<<"GGA";break;
				case 57:  outfile<<"GGT";break;
				case 43:  outfile<<"GGC";break;
				case 45:  outfile<<"GGG";break;
				case 65:  outfile<<"A";break;    //there's 'N's nearby
				case 84:  outfile<<"T";break;
				case 67:  outfile<<"C";break;
				case 71:  outfile<<"G";break;
			}
			//outfile.write(s_enter,j);
		}
		else   //case that "/ns/"
		{
            infile.get(ch);
			for(i=0,j=0;ch!='/';)
			{
                j=10*j;
				j+=(int)(ch)-48;   //accouting the number's of 'N"
                i++;
                infile.get(ch);
            }	
			for(i=1;i<=j;i++)
				outfile<<'N';
		}
	}

	infile.close();
	outfile.close();
	return 0;
}
\end{verbatim}

\section{Linux Shell that we used to test LZ77 performance}
\begin{verbatim}
#!/bin/sh
echo "please input a file name"
read file_name
date
num=1
while [ $num -lt 1000 ]
 do
gzip $file_name
gzip -d *.gz
num=$((num+1))
 done
date
\end{verbatim}
\end{document}